\documentclass[preprint,12pt]{elsarticle}
\usepackage{epsfig}
\usepackage{amssymb}

\journal{Nuclear Instruments and Methods in Physics Research Section A}

\begin{document}

\begin{frontmatter}

\title{A method for detection of muon induced electromagnetic showers with \mbox{the ANTARES detector}}

\author{J~A~Aguilar$^{1}$, I~Al~Samarai$^{2}$, A~Albert$^{3}$, M~Andr\'e$^{4}$, M~Anghinolfi$^{5}$, G~Anton$^{6}$, S~Anvar$^{7}$, M~Ardid$^{8}$, A~C~Assis~Jesus$^{9}$, T~Astraatmadja$^{9,a}$, J~J~Aubert$^{2}$
% ...
}

\address{$^1$ IFIC - Instituto de F\'isica Corpuscular, Edificios Investigaci\'on de Paterna, CSIC - Universitat de Val\`encia, Apdo. de Correos 22085, 46071 Valencia, Spain}
\address{$^2$ CPPM - Centre de Physique des Particules de Marseille, CNRS/IN2P3 et Universit\'e de la M\'editerran\'ee, 163 Avenue de Luminy, Case 902, 13288 Marseille Cedex 9, France}
\address{$^3$ GRPHE - Institut universitaire de technologie de Colmar, 34 rue du Grillenbreit BP 50568 - 68008 Colmar, France }
\address{$^4$ Technical University of Catalonia,Laboratory of Applied Bioacoustics,Rambla Exposició,08800 Vilanova i la Geltr\'u,Barcelona, Spain}
\address{$^5$ INFN - Sezione di Genova, Via Dodecaneso 33, 16146 Genova, Italy}
\address{$^6$ Friedrich-Alexander-Universit\"{a}t Erlangen-N\"{u}rnberg, Erlangen Centre for Astroparticle Physics, Erwin-Rommel-Str. 1, 91058 Erlangen, Germany}
\address{$^7$ Direction des Sciences de la Mati\`ere - Institut de recherche sur les lois fondamentales de l'Univers - Service d'Electronique des D\'etecteurs et d'Informatique, CEA Saclay, 91191 Gif-sur-Yvette Cedex, France}
\address{$^8$ Institut d'Investigaci\'o per a la Gesti\'o Integrada de Zones Costaneres (IGIC) - Universitat Polit\`ecnica de Val\`encia. C/  Paranimf 1, 46730 Gandia, Spain.}
\address{$^9$ Nikhef, Science Park,  Amsterdam, The Netherlands}
% ...
\address{$^{a}$ Also at University of Leiden, the Netherlands}
% ...

\begin{abstract}
The primary aim of ANTARES is neutrino astronomy with upward going muons
created in charged current muon neutrino
interactions in the detector and its surroundings.
Downward going muons are background
for neutrino searches. These muons 
are the decay products of cosmic-ray
collisions in the Earth's atmosphere far above the detector.
This paper presents a method to identify and count
electromagnetic showers induced along atmospheric muon tracks
with the ANTARES detector. The method is applied to both cosmic muon data and simulations
and its applicability to the reconstruction of muon event energies 
is demonstrated. 
\end{abstract}

\begin{keyword}
Neutrino telescope \sep Electromagnetic shower identification \sep High energy muons \sep Energy reconstruction.


\end{keyword}

\end{frontmatter}


\section{Introduction}
\label{sec:introduction}

The ANTARES neutrino telescope is located at a depth of 2475 m 
in the Mediterranean Sea, roughly \mbox{40 km} 
offshore from Toulon in France. 
Its main objective is the observation of 
extraterrestrial neutrinos. 
Relativistic charged leptons produced by neutrino interactions in 
and around the detector produce 
Cherenkov light in the sea water. 
This light is detected by an array of photomultiplier
tubes, allowing the muon direction to be reconstructed. 

Although the ANTARES detector is optimised for upward going particle
detection, the most abundant signal is due to atmospheric
downward going muons \cite{Coll2009,Coll2010,Coyle} produced 
in the particle showers induced by the
interactions of cosmic-rays in the atmosphere. 
In order to reduce this background the Earth is 
used as a filter, restricting the search
for cosmic neutrinos to sources in the Southern sky.

The processes contributing to the energy loss of a muon in water
include ionization, \mbox{$e^+e^-$ pair} production,
bremsstrahlung, and photonuclear 
interactions \cite{PDG,Groom,Klimush,Gaisser}. 
Below about  \mbox{1 TeV}, the muon 
energy loss is dominated by the continuous ionization process.
Above about  \mbox{1 TeV}, the muon energy loss 
is characterised by large energy fluctuations and 
discrete bursts. These bursts originate from 
pair production and bremsstrahlung (electromagnetic showers). 
The photonuclear interaction processes are less frequent and in the following 
no distinction is made between photonuclear induced showers 
and electromagnetic showers.
The average muon energy loss per unit track length due to 
these electromagnetic showers increases linearly with 
the energy of the muon \cite{PDG}.

A reconstruction algorithm is presented to 
identify electromagnetic showers induced by highly energetic muons 
with the ANTARES detector. The shower reconstruction algorithm relies 
on the identification of increased photon emission along the muon trajectory.
Counting electromagnetic showers along muon tracks to give an
estimate of the muon energy is called the \textit{pair meter} method \cite{Kokoulin, Gandhi}. 
The estimate of the energy of muons is important for many research 
topics. For example, an alternative method for estimating 
the energy of muons based on the occurrence rate of repeated measured photons 
on the photomultiplier tubes \cite{flux} has been used 
by the ANTARES Collaboration to search 
for a diffuse flux of cosmic high energy muon neutrinos.
Moreover, the angular resolution
of the muon trajectory
could benefit from a precise
discrimination of photon emission mechanisms along the
estimated track.
A similar measurement technique as the one presented in this 
article has been
published recently by the Super-Kamiokande Collaboration 
and used to select a sample of upward going muons with
energies above a TeV \cite{Desai}.  

\section{The ANTARES detector}                          
\label{detector}

A detailed description of the ANTARES detector 
is given elsewhere \cite{time,ARS,DAQ,Ant2}.

The full detector consists of twelve vertical
lines approximately \mbox{$450$ m} in height
equipped with a total of 885 photomultiplier tubes (PMTs).
The lines,
separated from each other by about \mbox{$65$ m}, are 
anchored to the sea floor by a dead weight and held taut by a buoy located 
at the top. The instrumented
part of the line starts \mbox{100 m} above the sea floor and consists 
of 25 floors with a separation distance of 
\mbox{14.5 m} along the line. The distance from the highest floor to the 
sea surface is around 2000 m. A floor consists of three 
PMTs pointing downward at an angle of $45^{\circ}$ with respect
to the vertical direction, in order to maximise the detection efficiency of upward going
tracks.

ANTARES is operating in the so called 
all-data-to-shore mode: all signals above a charge threshold (typically 0.3 photoelectrons) 
are digitised offshore
and sent to
shore to be processed in a computer farm. 
This farm applies a set of trigger criteria in order to separate muon-induced Cherenkov 
light from background light. The main sources of background light are the decay of $^{40}$K 
nuclei and the bioluminescence from organisms in the sea water.

\section{Algorithm for shower identification}                          
\label{algorithm}

The technique of the electromagnetic shower identification aims 
at distinguishing
Cherenkov photons emitted continuously along the muon track, hereafter 
called \textit{muon Cherenkov photons},
from the Cherenkov photons induced by electromagnetic showers, hereafter
called \textit{shower photons}.
Because of the short radiation length in water \mbox{($X_0=35$ cm)}, these 
showers rarely extend more than a few meters and can be considered point-like 
light sources for the ANTARES detector.
The electromagnetic showers are identified by an excess of photons 
above the continuous baseline of Cherenkov photons emitted by 
a minimum ionizing muon.  

The shower identification algorithm consists of two steps. 
The first step allows the identification and reconstruction of  
muon tracks. In the second step, a distinct shower candidate is identified by 
a cluster of measured photons at a particular point along the muon path.
The criteria to isolate this cluster are determined 
using a simulation code based 
on Corsika \cite{Corsika}. 
 
\subsection{Simulation}
\label{simulation}
Cosmic-ray interactions in the atmosphere, 
including atmospheric shower development, were
simulated with Corsika for primary energies between 
\mbox{1 TeV} and \mbox{$10^5$ TeV}, and incident angles between zero 
(vertical downward going) and 85 degrees.
The primary cosmic-ray composition and flux model employed 
is a simplified version of the H\"orandel parametrisation\footnote{The primary composition of the flux is subdivided into only five mass groups, namely proton, helium, nitrogen, magnesium and iron.} \cite{Hoerandel}. 
The chosen hadronic interaction 
model is QGSJET \cite{QGSJET}. The result of the Corsika simulation
is a set of muon tracks 
with their position, arrival time and momentum given at the surface of the sea. 
Typically, a single interaction leads to many muons at the sea surface. 
These muons are propagated through water. The discrete energy losses 
at high energies, the 
Cherenkov light production and propagation, including scattering,  
and the response of the detector
are simulated using a dedicated simulation package \cite{Km3_0, Bru}.
The muon propagation is performed by MUSIC 
\cite{MUSIC} in steps of 1 m. If
the energy loss of the muon over the step exceeds a given threshold \mbox{(1 GeV)},
an electromagnetic shower is simulated and shower photons are emitted. 
If the energy loss of
the muon over the step
is below the threshold, only muon Cherenkov photons are simulated. 
The simulation package also uses tables generated from 
\mbox{GEANT 3} \cite{GEANT} that 
parameterise the arrival time 
and the amount of light detected by individual PMTs. These tables 
take into account the measured properties of the water at the ANTARES site, 
the angular dependence of the acceptance of the PMT and also the position,
distance and orientation of the PMT with respect to a given muon track.
The optical background is assumed to be constant at a rate of 
60 kHz \cite{Escoffier} on each PMT.

\subsection{Algorithm}
\label{algorithm1}

\begin{figure}[tb]
 \setlength{\unitlength}{1cm}
 \centering
 \includegraphics[width=13.5cm]{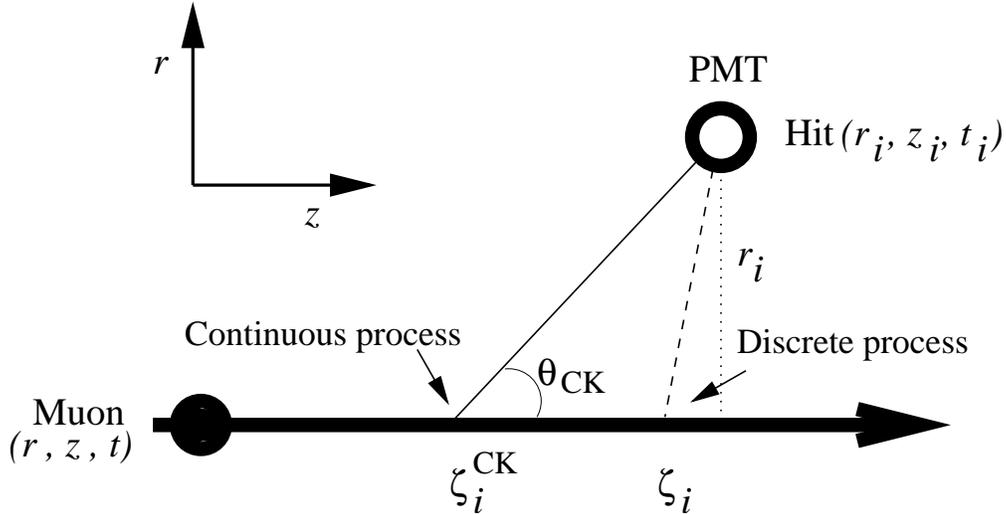}
 \caption[]{Schematic view of muon Cherenkov light detection. The thick line represents the muon trajectory, the thin line the path of Cherenkov light and the thin dashed line the path of shower light. The muon goes through a reference point $(r, z, t)$. The Cherenkov light is emitted at an angle $\theta_{\rm CK}$ with respect to the muon track at point $\zeta_i^{\rm CK}$ and is detected by a PMT as a hit at point $(r_i, z_i, t_i)$. The shower light is emitted at point $\zeta_i$ and is detected by the same PMT at a different time.}
\label{fig:calc}
\end{figure}

The shower identification algorithm proceeds in several steps. 
First, the muon trajectory must be determined. This is done using a 
standard tracking algorithm \cite{line1, Ronald} that provides an estimate of 
the direction and position of the muon at a given time.
In what follows, a hit is taken to be a photomultiplier signal exceeding a charge threshold of 0.3 photoelectrons~\cite{PMT}.
Using the configuration in Figure \ref{fig:calc}, the expected Cherenkov photon arrival time $t^{\rm CK}_i$ for each hit $i$ is calculated as
\begin{equation}
t^{\rm CK}_i = t + \frac{1}{c} \Big(z_i-z - \frac {r_i}{\tan \theta_{\rm CK}}\Big) + \frac{n}{c} \frac{r_i}{\sin \theta_{\rm CK}},
\label{Eq:CKlight}
\end{equation}
where $t$ is the time at which the muon passes point $(r,z)$, $\frac{c}{n}$ is the group velocity of light 
in water ($n$ = 1.38 is the group refractive index 
for ANTARES water), $\theta_{\rm CK}$ is the Cherenkov angle for a relativistic muon in water
($\theta_{\rm CK}\sim42^o$)
and $r_i$ is the perpendicular distance between the 
muon trajectory and the PMT. 

\begin{figure}[tb]
 \setlength{\unitlength}{1cm}
 \centering
 \includegraphics[width=15.0cm]{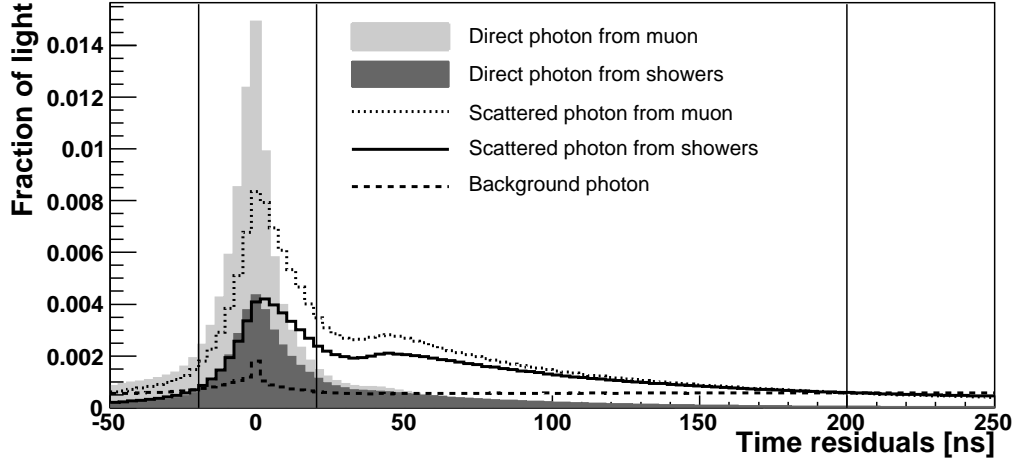}
\caption[Sc]{Time residuals for the measured photon arrival times relative to the calculated arrival times of Cherenkov photons coming from reconstructed muon tracks in a Monte Carlo sample. Contributions are shown for the direct and scattered photons originating from the muon as well as from the showers. Also shown are background photons. The three vertical lines define the early time interval (between -20 ns and 20 ns)  and the late time interval (between 20 ns and 200 ns). The enhancement at \mbox{45 ns} is due to an effect of the PMT read-out electronics, which has been included in the simulation.} 
\label{fig:timeresidual}
\end{figure}

The fitted trajectory can be used to 
characterize hits by their arrival times into 
three groups: early hits that are 
predominantly due to Cherenkov photons, late hits that are 
mainly due to scattered and shower photons, and extremely 
early or late hits that can safely be assumed to be due to background.
Figure \ref{fig:timeresidual}
shows time residuals ($t_i-t^{\rm CK}_i$) for muon energies between 100 GeV 
and 100 TeV generated by the simulation
described in Section \ref{simulation}.
Direct hits have a roughly Gaussian  
distribution (with a long tail of late hits) with a peak 
at zero time residual and a full width at half maximum of $\sim$20 ns.

Early hits ($|t_i-t^{\rm CK}_i| < t_{min}, t_{min}= 20$ ns)  
contain mostly muon Cherenkov photons whose emission positions 
along the muon track are given by
\begin{equation}
\zeta_i^{\rm CK}=z_i- z - \frac {r_i}{\tan \theta_{\rm CK}}.
\label{equ:cvpos}
\end{equation}
The variation in the arrival time of these Cherenkov hits can 
be attributed to the dispersion of light in the sea water. 
Note that Equation \ref{equ:cvpos} is used to determine 
the emission point of all photons leading to early hits, even shower 
photons that may not be emitted at the Cherenkov angle.

Late hits ($t_{min} < t_i-t^{\rm CK}_i < t_{max}, t_{max}= 200$ ns) 
contain the 
largest fraction of hits due to shower photons. The value of $t_{max}$ 
has been taken to be the point at which a hit is equally likely to be 
due to a shower photon as to a background photon. 
These shower photons may not necessarily be emitted at the Cherenkov angle 
from the muon track. Therefore the emission angle is left as a free parameter 
and, with the photon emission taking 
place at $\zeta_i$ (see Figure \ref{fig:calc}), 
the hit time is given by 
\begin{eqnarray}
t_i = t + \frac{\zeta_i-z}{c} + \frac{n}{c} \sqrt{ r_i^2 + (z_i- \zeta_i )^2 }.
\label{equ:showertime}
\end{eqnarray}
Equation (\ref{equ:showertime}) has two distinct 
solutions, $\zeta_i^{+}$ and $\zeta_i^{-}$.

Extremely late or early hits ($t_i > t_{max}$ or $t_i <-t_{min}$) 
are assumed to be background hits and are rejected.

All calculated $\zeta_i^{\rm CK}$, $\zeta_i^{+}$ and $\zeta_i^{-}$  positions 
along the muon track are collected in a one dimensional histogram. 
The shower position is identified by the localised increase of 
the number of emitted photons along the reconstructed muon trajectory, 
identified by a peak in the histogram.
If the two solutions $\zeta_i^{+}$ and $\zeta_i^{-}$ are found in different peaks,
the shower identification procedure will ignore the solution in the smaller peak.

\begin{figure}[]
 \setlength{\unitlength}{1cm}
 \centering
 \begin{picture}(15.5,14.5)
\put(-1.0,0.0){\epsfig{file=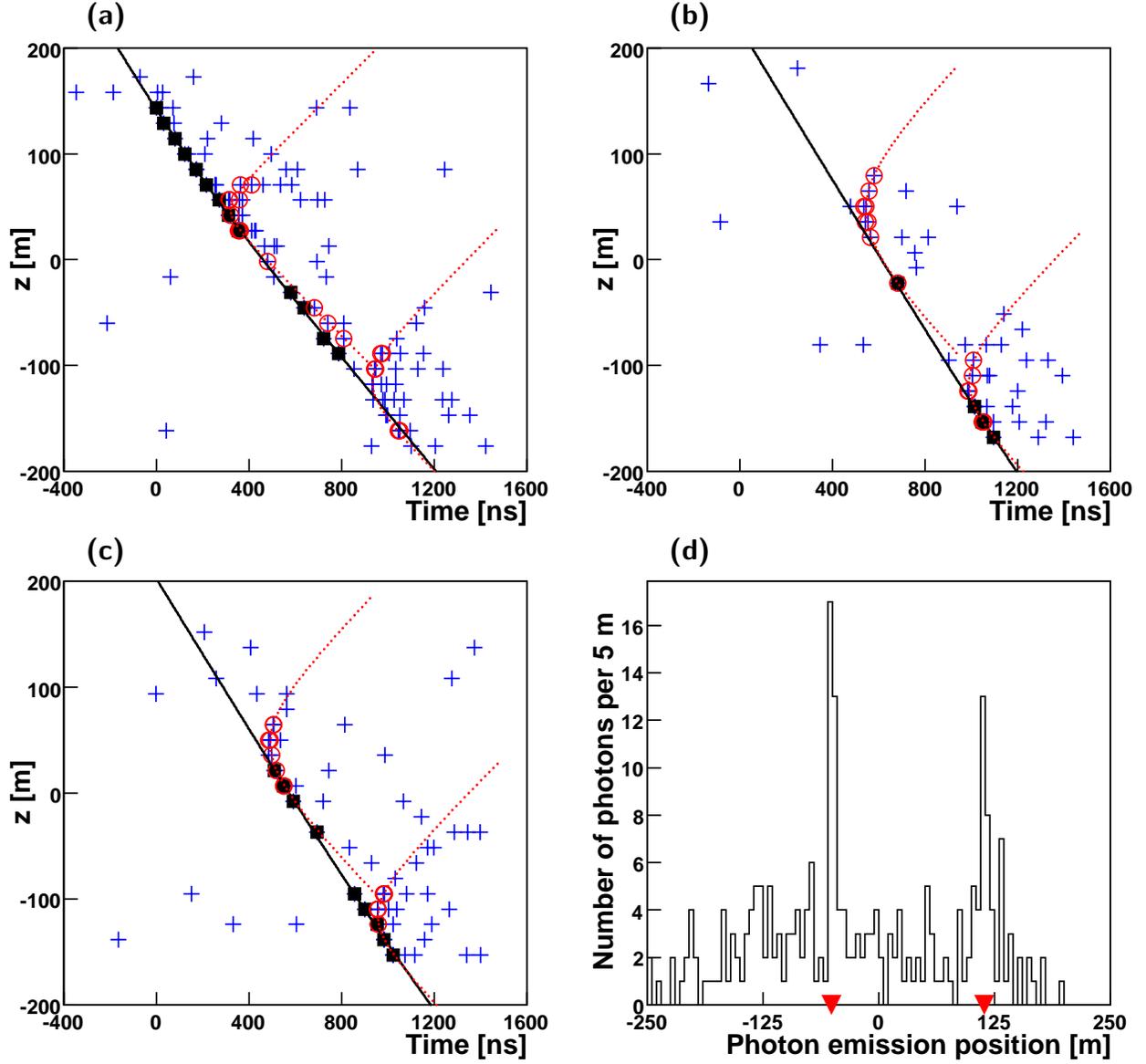,width=17.0cm,clip=}}
   \put(0.2,15.1){\textbf{\textsf{(a)}}}
   \put(8.7,15.1){\textbf{\textsf{(b)}}}
   \put(0.2,7.3){\textbf{\textsf{(c)}}}
   \put(8.7,7.3){\textbf{\textsf{(d)}}}
\end{picture}
\caption[Sc]{Display of an atmospheric muon event. The first three panels (a)(b)(c) show, for each line, the altitude of the photomultiplier tube for each associated photon (crosses) as a function of the arrival time of the photon. The origin on the z-axis corresponds to the middle of the line and the time offset is chosen with respect to the time of the first detected photon compatible with the muon trajectory. The muon track (solid line) and two electromagnetic \mbox{showers} (curved dotted lines) are reconstructed. The black squares indicate identified photons which are used in the muon trajectory reconstruction. The empty circles around the crosses indicate photons used in the shower reconstruction. The bottom right plot (d) shows the number of detected photons projected along the muon trajectory. The peaks correspond to the reconstructed shower positions indicated by the triangles.}
\label{fig:EvDisplay}
\end{figure}

An example of the application of this procedure to data can be seen in Figure \ref{fig:EvDisplay}. The bottom right panel of this figure 
shows the emission points of photons along the muon trajectory, as determined by the 
solutions of Equation \ref{equ:cvpos} and Equation \ref{equ:showertime}.
Two excesses are visible and are attributed to two electromagnetic showers.
Each of the other three panels shows a height versus time diagram of data obtained from 
one of the detector lines. 
The result of the muon trajectory reconstruction for a relativistic 
muon in water together with results of the shower 
identification are also displayed. 
A downward going muon with 
two electromagnetic showers is thereby identified.
Using the fitted shower positions from the shower identification
algorithm, a prediction is made for the arrival time of the shower light.
The dotted curves in Figure \ref{fig:EvDisplay} show the expected photon arrival time under the assumption 
of a spherical light emission from the reconstructed position of the shower. 
As can be seen from \mbox{Figure \ref{fig:EvDisplay}}, most, but not all, photons
are associated with the muon track fit. Many photons that do not
comply with the muon track fit are associated to shower photons.
The photons not associated with the muon track 
or associated showers can be attributed to random 
background photons due to radioactive $^{40}\mathrm{K}$ decays  
and bioluminescence.

\section{Selection and performance}
\label{simulationandselection}

The selection and performance of the shower identification 
algorithm has been studied 
with the simulation described in Section \ref{simulation}.

\subsection{Muon and shower selection}
\label{sec:showermuonselect}

The shower identification algorithm is applied to 
well reconstructed muon tracks that have the potential 
to produce a detectable shower. Two criteria have been used to select 
such tracks. First, the track length is required to be at 
least 125 m. The track length is taken to be the distance 
between the emission points of the photons that gave rise to 
the earliest and latest hit used in the muon track 
reconstruction. Second, the muon trajectory reconstruction 
is required to have used at least 12 hits. 
These criteria select about 65\% of 
all reconstructed (downward going) muon tracks.

The shower-induced photon emission along the muon
trajectory results in localised peaks as shown in Figure \ref{fig:EvDisplay}d.
The task to identify a shower is then reduced to a one dimensional
peak finding algorithm whose result can be characterised by 
three parameters, namely the center, width and height of the
identified peak.
Potential peaks are identified through the subtraction of 
the constant photon background, as determined by a sensitive 
nonlinear peak clipping algorithm. This algorithm tracks the baseline 
of a spectrum by comparing the value of each data point with 
the average value of neighboring data points, taking the baseline 
to be the smaller value. Further details can be found in \cite{Morhac}.

For each potential peak, the number of hits is integrated
in a $\pm 5$ m window around the peak center. 
All hits are assumed to be single photons. 
Only peaks 
having at least 10 hits above Cherenkov photon baseline 
in this window of 10 m are selected (typically yielding peaks with three sigma significance).
The number of baseline hits is defined as the 
average number of hits along the track times the window of 10 m.
In order to suppress wrongly identified showers, 
hits from at least 
five different floors 
are required for each peak.

\subsection{Performance of the shower identification method}
\label{sec:performance} 

\begin{figure}[tb]
 \setlength{\unitlength}{1cm}
 \centering
 \includegraphics[width=8.5cm]{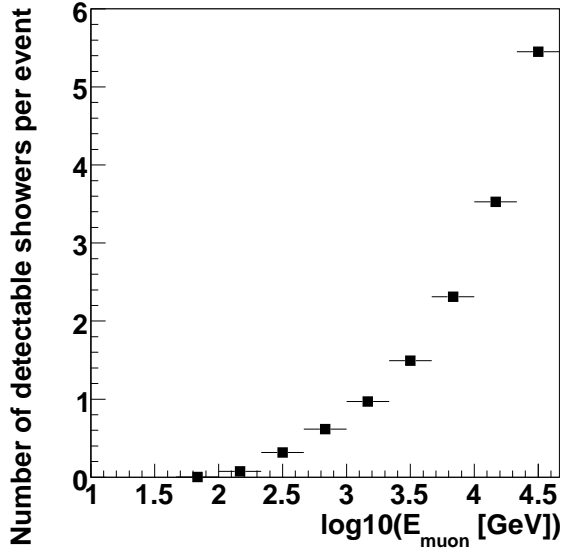}
 \caption[Sc]{Average number of detectable showers which have shower photons detected on at least five different floors per atmospheric muon event as a function of the muon energy.}
\label{fig:showervsenergy}
\end{figure}

Figure \ref{fig:showervsenergy} shows the number of detectable showers,
coming from well reconstructed muons,
that have shower photons detected on at least five different floors 
per atmospheric muon event as a function
of the muon energy. 
The atmospheric muon events are usually muons in a 
bundle with an average multiplicity around 3.3~\cite{Corsika}. 
The muon energy in Figure \ref{fig:showervsenergy} refers 
to the muon with the largest energy in the bundle.
These muons have an average energy of \mbox{1.2 TeV}
and their mean generated shower energy is around 120 GeV.
Muons with at least one identified shower
have on average 2.5 times higher energy than 
muons without an identified shower. 

The event selection and algorithm has been tuned to count the number 
of showers with a high level of purity,
even at the expense of efficiency. 
In order to study the efficiency and purity of the reconstruction, 
the Monte Carlo truth information is consulted to determine 
whether the result of the shower reconstruction corresponds 
to any actual shower and, if so, whether that shower has been well 
reconstructed. A reconstructed shower is said to have correctly 
identified an underlying shower if its position is determined to 
within 25 m of the true generation point and if 25\% of the hits in the 
reconstructed shower are truly due to photons produced by the underlying 
shower. Here, a hit is in the reconstructed shower if its projected emission 
point along the muon track is within 5 m of the reconstructed shower 
position (see Figure \ref{fig:EvDisplay}d).

\begin{figure}[tb]
 \setlength{\unitlength}{1cm}
 \centering
 \begin{picture}(15.5,8.5)
   \put(-1.5,0.3){\epsfig{file=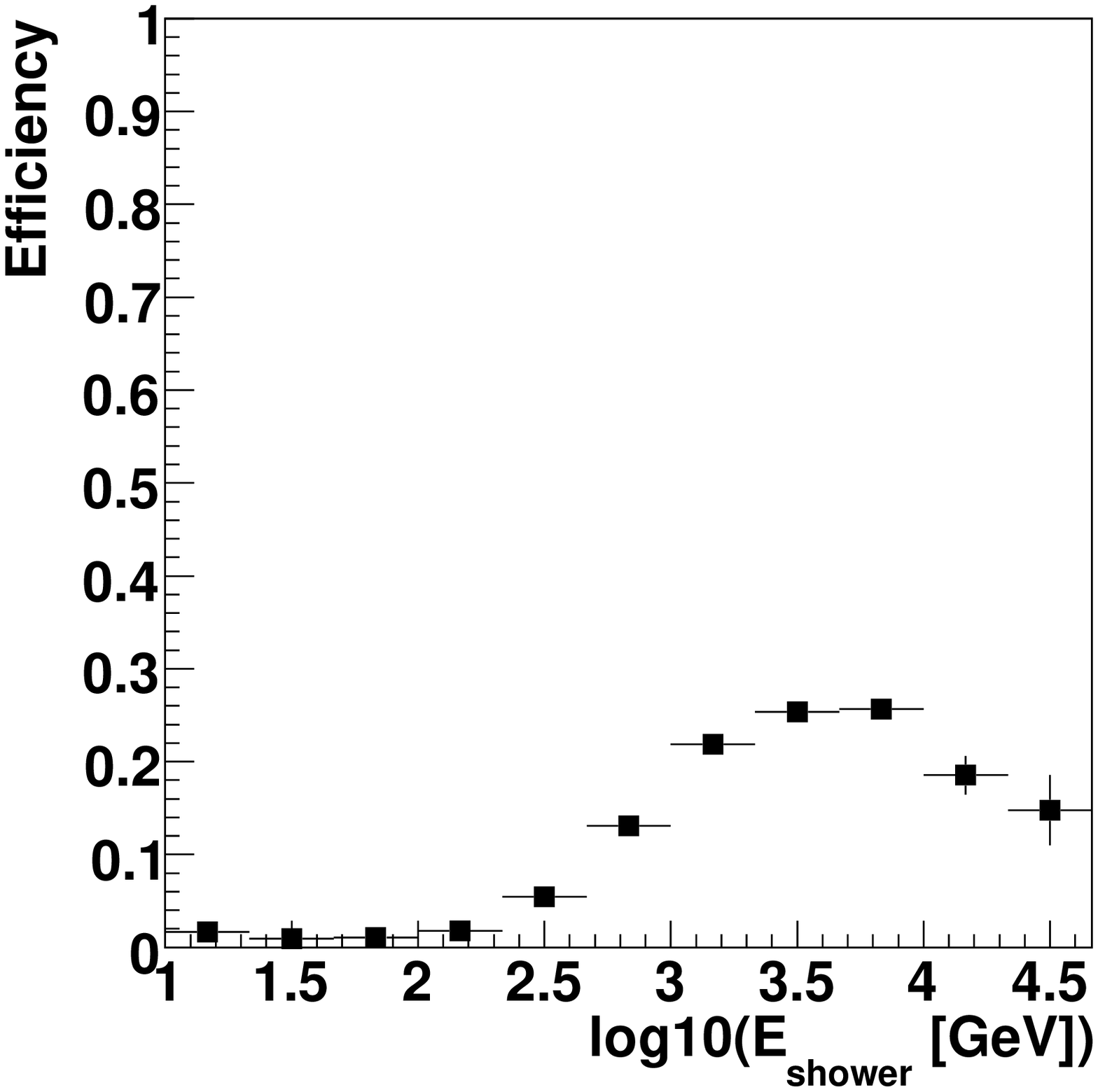,width=8.5cm,clip=}}
   \put(6.5,0.3){\epsfig{file=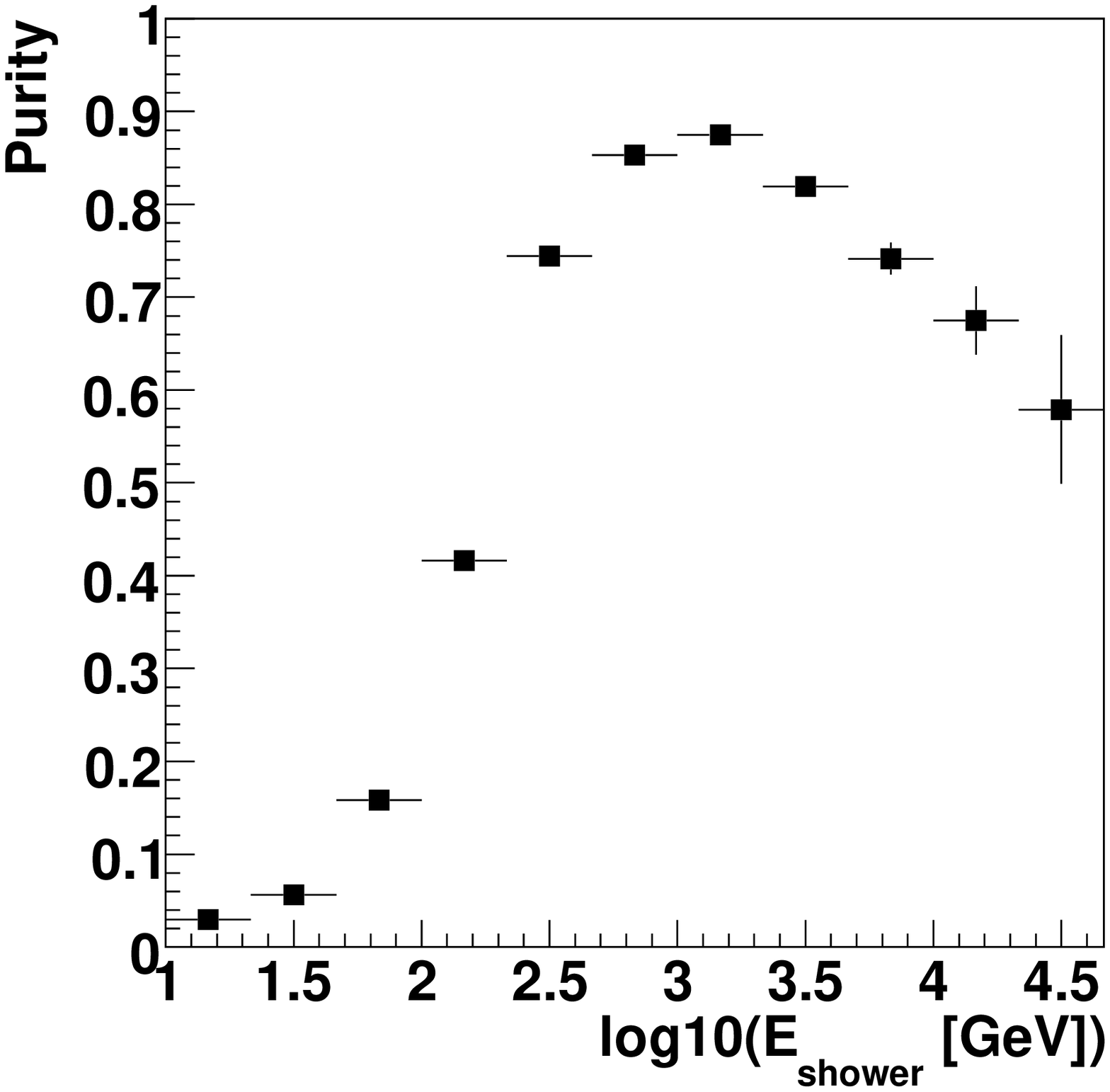,width=8.5cm,clip=}}
  \put(0.1,7.1){\textbf{\textsf{(a)}}}
  \put(8.1,7.1){\textbf{\textsf{(b)}}}
 \end{picture}
\caption[Sc]{(a) Efficiency and (b) purity as a function of the shower energy for reconstructed showers obtained with a Monte Carlo sample of atmospheric muons.}
\label{fig:showereff}
\end{figure}

The efficiency with which showers are correctly identified is given 
by the ratio of the number of well identified showers to the total 
number of simulated showers that give rise to hits in
at least five different floors, i.e. 
all showers that may reasonably be expected to be reconstructed. 
The shower identification efficiency ranges from 5\% at low shower energy 
($\sim$300 GeV) to 30\% at high shower energy ($\sim$5 TeV), as shown in Figure \ref{fig:showereff}a.

The purity of the reconstructed shower sample is given by the 
ratio of the number of correctly identified showers to the number 
of all reconstructed showers and is shown in Figure \ref{fig:showereff}b. 
The shower purity ranges from 40\% at low 
shower energy ($\sim$300 GeV) to 90\% at high shower energy ($\sim$1 TeV). At 
even higher shower energies, the purity decreases, reaching 60\% at 30 TeV. 
Such showers are mainly produced by very high energy muons. The 
density of photons along the trajectory of such a muon is 
great enough that an excess of photons due to a particular shower becomes 
difficult to observe.

\section{Comparison between data and Monte Carlo simulations}

A sample of data corresponding to 47.3 days of data 
taking between January and December 2007 has been used to study the behavior of the shower identification algorithm.
During this period the detector comprised five lines.

\begin{figure}[tb]
 \setlength{\unitlength}{1cm}
 \centering
 \begin{picture}(15.5,8.5)
   \put(-1.5,0.3){\epsfig{file=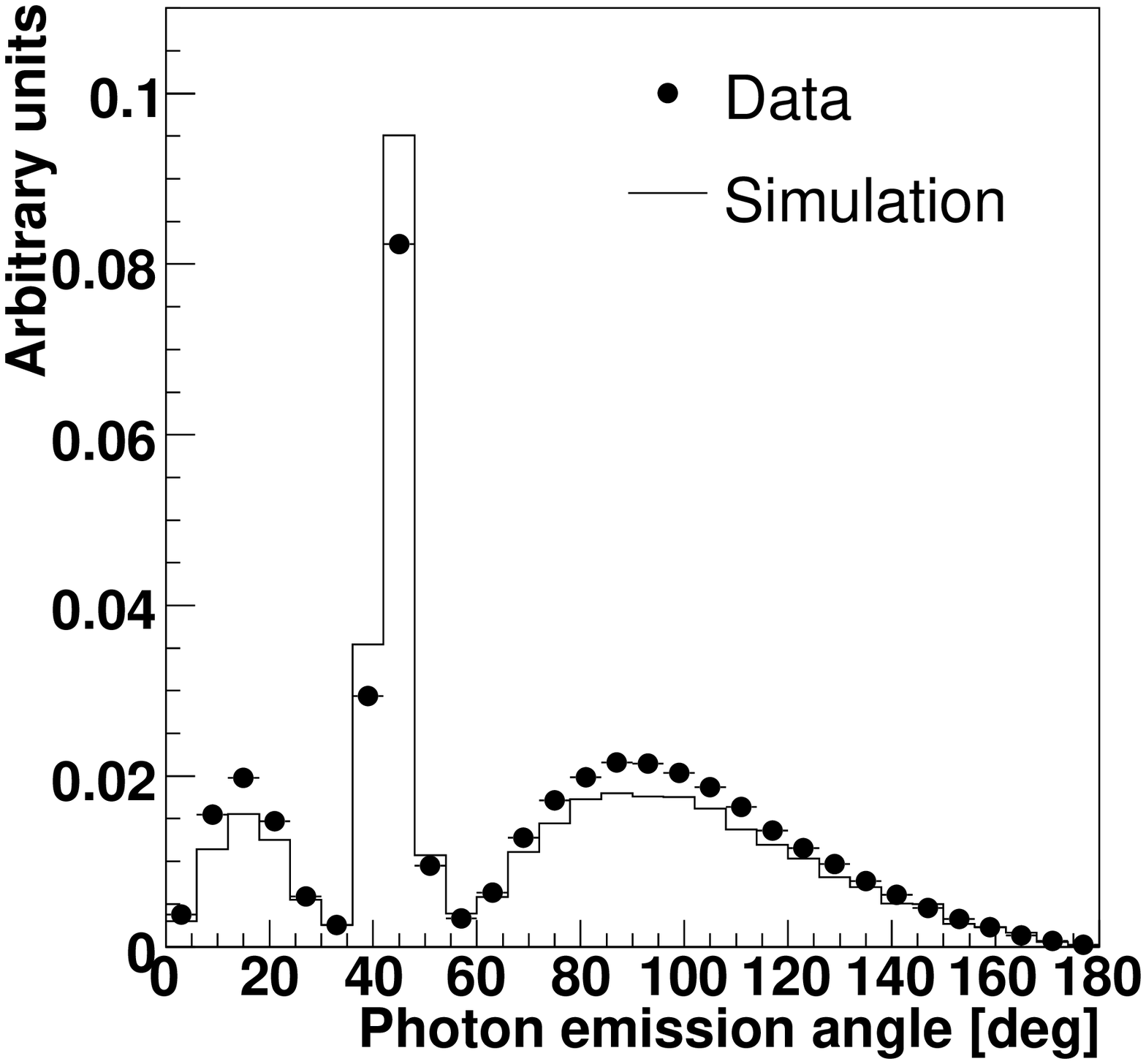,width=8.5cm,clip=}}
   \put(6.5,0.3){\epsfig{file=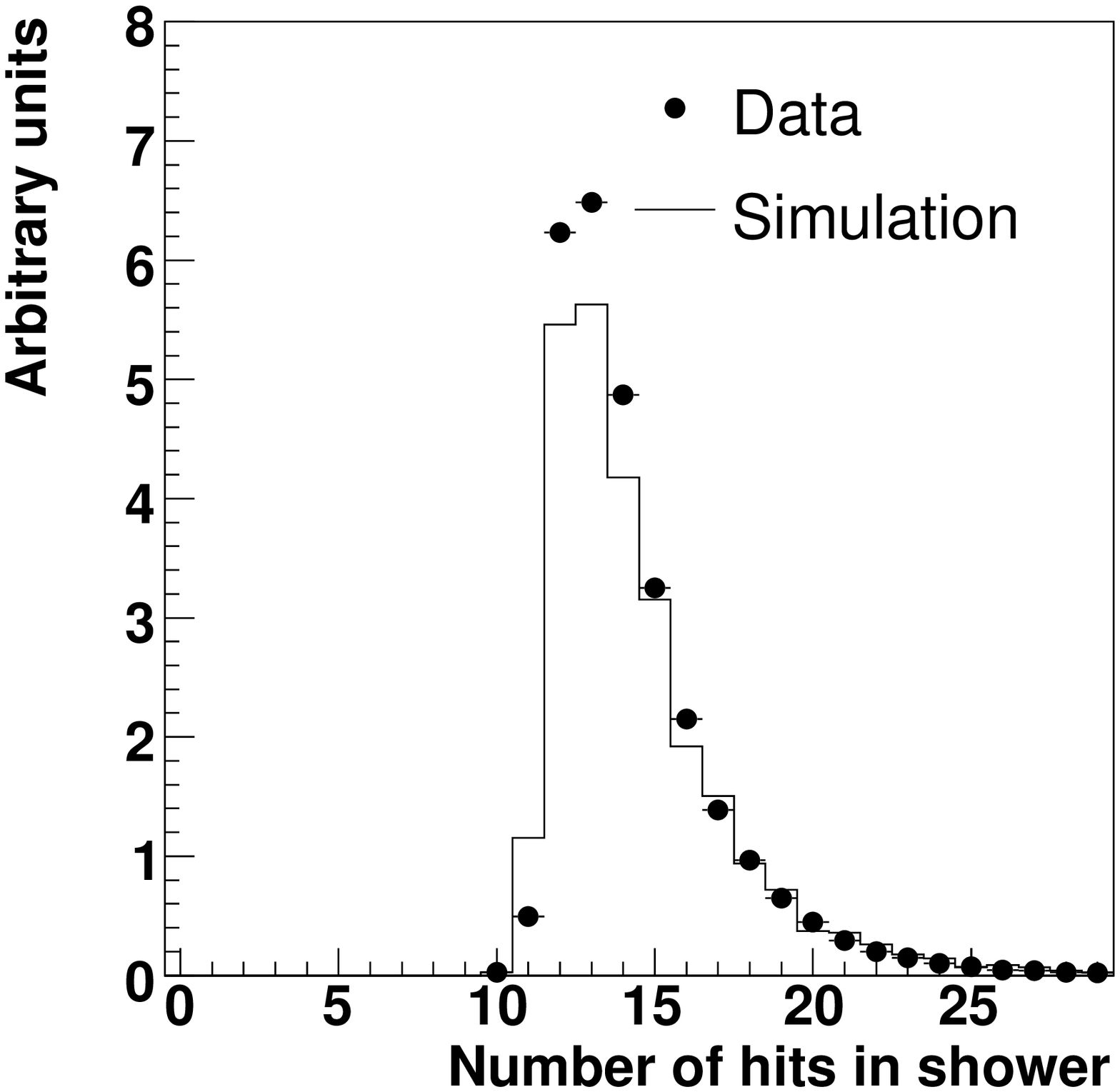,width=8.5cm,clip=}}
  \put(0.1,7.1){\textbf{\textsf{(a)}}}
  \put(8.1,7.1){\textbf{\textsf{(b)}}}
 \end{picture}
\caption[Sc]{(a) Photon emission angle and (b) relative number of hits (note that a minimal number of ten hits is required by the identification algorithm) used in the shower identification for the data and the Corsika simulation.}
\label{fig:showers1}
\end{figure}

The Corsika simulation (including the simplified H\"orandel flux) was scaled by a factor 0.83 to normalise 
the simulated muon rate to the measured muon rate for the selected tracks.
Figure \ref{fig:showers1}a shows 
the angular distribution of the
shower photons with respect to the muon direction. 
The shape of the distribution is determined  
by detector effects and the cuts used in the analysis. The peak 
around 42 degrees comes from shower photons emitted 
at the Cherenkov angle through showers oriented in the direction of the muon
and whose emission points have been calculated using Equation \ref{equ:cvpos}, whereas the other hits have been calculated using Equation \ref{equ:showertime}.

Figure \ref{fig:showers1}b shows the hit multiplicity distribution of  
selected showers. With the final set of cuts applied, 
the average number of hits associated to an identified shower is around 14. 
As the quantity of Cherenkov light produced by an electromagnetic shower increases linearly with the muon energy, 
counting the number of hits in one shower provides a first order estimate of its energy. 
The data distributions agree reasonably 
well with the Corsika simulation. 

\begin{table}[tb]
\begin{center}
\begin{tabular}{|l|r|r|r|r|r|r|}
\hline
Number of identified showers        & 0  & 1   & 2  & 3+  \\ 
\hline
Source of uncertainty                  & \multicolumn{4}{|c|}{Variation in [\%]}         \\                                   
\hline
Background rate        & 0.1  &  1  &  12 &   14  \\ 
Minimal shower energy  & 0.1  &  3  &   6 &   5  \\
PMT angular acceptance & 1.2  &  18 &  30 &   3   \\
Absorption length      & 1.3  &  17 &  39 &   77  \\
\hline
Total systematic uncertainty [\%]  & $\pm$ 1.8  & $\pm$ 25  & $\pm$ 51 & $\pm$ 78  \\  
\hline
\end{tabular}
\end{center}
\caption[]{Variation in the number of identified showers as the values of selected Monte Carlo parameters are changed. The systematic uncertainty is estimated by varying the background rate, the energy threshold to produce photons from electromagnetic shower light, the PMT angular acceptance and the water absorption length (see text).}
\label{tab:systematic}
\end{table}

The simulation has also been used to evaluate
systematic uncertainty on the number of identified showers.
\mbox{Table \ref{tab:systematic}} shows the systematic uncertainty 
determined by varying parameters in the simulation \cite{Coll2010}. 
The measured detector 
background rate is around 60 kHz on average \cite{Escoffier}.
The systematic error arising from the uncertainty of this background rate is estimated by 
repeating the analysis with a background rate of 50 kHz and with a background rate of 120 kHz (row three in Table \ref{tab:systematic}).
The values are the percentage variation with respect to the values from the default simulation.
Uncertainties arising from the
energy threshold to produce hits from electromagnetic showers  
or hits from muon Cherenkov light
is estimated by varying the threshold $\pm 50$\% from its default value
of \mbox{1 GeV} (row four in Table \ref{tab:systematic}). 
Uncertainty on the angular acceptance of the optical modules is estimated 
by taking a different parametrization of the PMT angular acceptance \cite{Coll2010} (row five in Table \ref{tab:systematic}).
The water properties 
are taken into account by varying the 
absorption length by $\pm 20$\% around the measured value \cite{transmission} (row six in Table \ref{tab:systematic}).

All systematic uncertainties are added in quadrature. 
The largest contributions to the systematic error arise from uncertainties
on the PMT angular acceptance and on the water absorption length.
When decreasing 
the absorption length, fewer showers are identified, since more photons are absorbed in the water 
before they reach the PMTs. On the other hand, the systematic studies 
show evidence that the shower algorithm 
is robust against large variations of the background, because showers emit
a light density much bigger than that of the optical background. 

\begin{figure}[tb]
 \setlength{\unitlength}{1cm}
 \centering
\includegraphics[width=8.5cm]{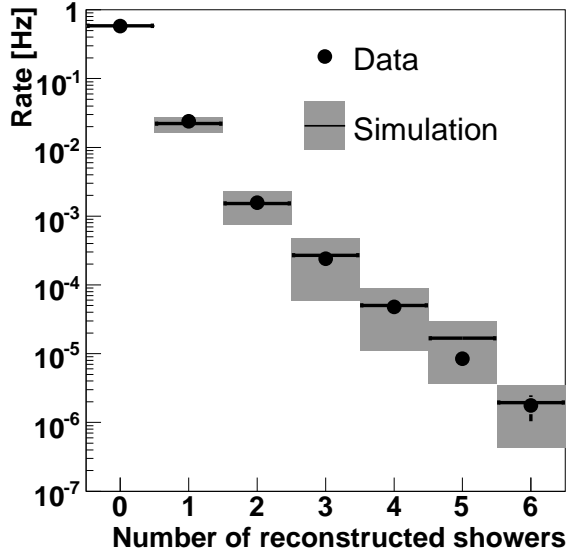} 
\caption[Sc]{Muon event rate as a function of the shower multiplicity for data (points) and the Corsika simulation (line) with no correction for the identification efficiency. The systematic error for the simulation is given by the height of the grey bands. Only statistical errors are shown for the data points.}
\label{fig:showers}
\end{figure}

The muon event rate as a function of the number of identified 
showers is shown 
in \mbox{Figure \ref{fig:showers}}. 
The distribution shows
the results for data and the Corsika based simulation with no correction for the 
identification efficiency.
Also shown is the systematic uncertainty for the simulation. 
For the data points, only the statistical errors
are shown. 
As can be seen, about 4\% of the 
selected muon tracks have one well identified shower.
There is agreement between data and Monte Carlo over five orders of magnitude.

\section{Conclusions}
\label{conclusion} 

A method to identify electromagnetic showers emitted by
muons has been developed, characterised and 
applied to the downward going muon data taken with
the ANTARES detector.
This algorithm exploits the different emission characteristics
of shower-induced and primary muon-induced Cherenkov photons.
The shower light emission is localised at discrete
points along the muon trajectory, whereas the traversing muon continuously
emits Cherenkov photons under a constant and known angle relative to the muon
trajectory.
The essential element of the algorithm is 
the projection and identification of photon vertices along the muon
track with a subsequent peak finding algorithm.
The performance of the identification
algorithm has been validated using a sample of simulated
atmospheric muon events and agreement was found in the number of identified showers 
between data and simulations. 

With the development and validation of this electromagnetic shower multiplicity
estimator, important new information becomes available
for physics analysis. In particular the method establishes a first step 
towards a new energy estimator.
With the application of this method, it can be concluded that stochastic 
energy loss has been observed in ANTARES.

\section*{Acknowledgments}

The authors acknowledge the financial support of the funding agencies:
Centre National de la Recherche Scientifique (CNRS), Commissariat
\'a l'\'ene\-gie atomique et aux \'energies alternatives  (CEA), Agence
National de la Recherche (ANR), Commission Europ\'eenne (FEDER fund
and Marie Curie Program), R\'egion Alsace (contrat CPER), R\'egion
Provence-Alpes-C\^ote d'Azur, D\'e\-par\-tement du Var and Ville de
La Seyne-sur-Mer, France; Bundesministerium f\"ur Bildung und Forschung
(BMBF), Germany; Istituto Nazionale di Fisica Nucleare (INFN), Italy;
Stichting voor Fundamenteel Onderzoek der Materie (FOM), Nederlandse
organisatie voor Wetenschappelijk Onderzoek (NWO), the Netherlands;
Council of the President of the Russian Federation for young scientists
and leading scientific schools supporting grants, Russia; National
Authority for Scientific Research (ANCS), Romania; Ministerio de Ciencia
e Innovaci\'on (MICINN), Prometeo of Generalitat Valenciana and MultiDark,
Spain. We also acknowledge the technical support of Ifremer, AIM and
Foselev Marine for the sea operation and the CC-IN2P3 for the computing facilities.

\bibliographystyle{report}  
\bibliography{sabibnew}		

\end{document}